\documentclass{Interspeech2024}
\usepackage{multirow}
\usepackage{xcolor}
\usepackage{hyperref}       % hyperlinks
\usepackage{url}            % simple URL typesetting

\interspeechcameraready 

\title{SimulTron: On-Device Simultaneous Speech to Speech Translation}

\name[affiliation={1}]{Alex}{Agranovich}
\name[affiliation={1}]{Eliya}{Nachmani}
\name[affiliation={1}]{Oleg}{Rybakov}
\name[affiliation={2}]{Yifan}{Ding}
\name[affiliation={2}]{Ye}{Jia}
\name[affiliation={1}]{Nadav}{Bar}
\name[affiliation={2}]{Heiga}{Zen}
\name[affiliation={1}]{Michelle}{Tadmor Ramanovich}

\address{
  $^1$Google Research
  $^2$Google DeepMind}
\email{alexagr@google.com, eliyn@google.com}


\begin{document}

\maketitle

\begin{abstract}
    Simultaneous speech-to-speech translation (S2ST) holds the promise of breaking down communication barriers and enabling fluid conversations across languages. However, achieving accurate, real-time translation through mobile devices remains a major challenge. We introduce SimulTron, a novel S2ST architecture designed to tackle this task. SimulTron is a lightweight direct S2ST model that uses the strengths of the Translatotron framework while incorporating key modifications for streaming operation, and an adjustable fixed delay. Our experiments show that SimulTron surpasses Translatotron 2 in offline evaluations. Furthermore, real-time evaluations reveal that SimulTron improves upon the performance achieved by Translatotron 1. Additionally, SimulTron achieves superior BLEU scores and latency compared to previous real-time S2ST method on the MuST-C dataset. Significantly, we have successfully deployed SimulTron on a Pixel 7 Pro device, show its potential for simultaneous S2ST on-device.
\end{abstract}

\section{Introduction}

Speech-to-speech translation (S2ST) is a transformative technology with the potential to break down language barriers and foster global connections. 
In recent years, groundbreaking models \cite{jia2019direct,zhang2021uwspeech,lee2021direct,lee2021textless,huang2022transpeech} have revolutionized the field of S2ST. These models have achieved remarkable performance, surpassing the traditional cascade-based approach with their direct S2ST translation methods. Moreover, they preserve speaker identity, intonation, and other subtle nuances that lend authenticity and expressiveness to speech. While S2ST technology continues to evolve, the challenge of real-time, on-device simultaneous translation persists. Existing simultaneous translation models \cite{sudoh2020simultaneous,ma2021direct,liu2021start,huang2022transpeech,barrault2023seamless,dugan2023learning} are not optimized for the inherent constraints of mobile devices. Today, with smartphones and tablets being central hubs for personal and professional interactions, on-device S2ST is crucial. This approach offers increased accessibility, privacy, and the ability to bridge linguistic divides.

We introduce SimulTron, a novel on-device, simultaneous S2ST model built upon the foundation of the Translatotron architecture \cite{jia2019direct}. SimulTron leverages the strengths of Translatotron while incorporating key modifications, including a causal conformer encoder, wait-k attention, convolutional post-net network, and streaming vocoder specifically tailored for the on-device, simultaneous translation scenario. Our contributions in this paper are threefold:
\begin{itemize}
    \item \textbf{Real-time On-Device Simultaneous S2ST:} We present on-device, simultaneous S2ST model, the model starts outputting the translation after an adjustable fixed delay for context. Paving the way for real-time, language-agnostic communication on mobile devices. SimulTron processes 320-sample audio packets in streaming mode. Features are extracted via mel spectrograms and encoded by a 16-layer causal conformer encoder. The encoded output is decoded in streaming mode and synthesized into audio by a streaming vocoder, enabling instant translation output during speech.
    \item \textbf{Enhanced Translatotron Architecture:} We propose improvements to the Translatotron architecture, leading to superior performance for both offline and simultaneous S2ST. Notably, our model demonstrates a clear improvement over the offline Translatotron architecture in simultaneous settings and surpasses the Translatotron 2 architecture in offline evaluations. Additionally, SimulTron achieves superior BLEU scores and latency compared to previous real-time S2ST methods on the MuST-C dataset. This demonstrates that SimulTron can effectively translate speech while preserving its natural characteristics, even under the constraints of on-device processing.
    \item \textbf{Comprehensive Analysis:} We conduct a rigorous evaluation of SimulTron, examining its performance across various floating-point precision levels and latency metrics. We also compare SimulTron's performance against offline models, providing valuable insights into its efficiency and effectiveness. This analysis sheds light on the trade-offs inherent in on-device S2ST translation.
\end{itemize}

\section{Related Work}
\subsection{Offline speech-to-speech translation}
Speech-to-speech translation has revolutionized cross-language communication, evolving from modular systems to end-to-end models. Traditionally, S2ST relied on a cascade of speech-to-text, machine translation, and text-to-speech systems \cite{nakamura2006atr,wahlster2013verbmobil,lavie1997janus}. However, the field has seen a significant shift towards end-to-end training models, offering increased efficiency and accuracy. The Translatotron \cite{jia2019direct} architecture pioneered direct S2ST, preserving speaker identity and using phoneme recognition loss, but initially lagged behind cascade systems in performance. Translatotron 2 \cite{jia2022translatotron} enhanced the original model with a linguistic decoder and a unique method for retaining speaker voice, significantly boosting its performance to rival cascade systems. Translatotron 3 \cite{nachmani2023translatotron} utilizes unsupervised learning with back-translation, enabling training on readily available speech data without requiring paired translations. Underscoring the importance of non-linguistic cues, Translatotron 3 uniquely emphasizes the preservation of non-linguistic cues in speech. Other notable works in unsupervised S2ST include \cite{wang2022simple, fu2023improving}, which achieve comparable results to supervised methods. Another significant development in S2ST involves using speech tokens for training models. Notable examples include UWSpeech \cite{zhang2021uwspeech}, Speech-to-Unit Translation (S2UT) \cite{lee2021direct,lee2021textless}, and Transpeech \cite{huang2022transpeech}. This approach is crucial because speech token-based S2ST models can be trained on unwritten languages, expanding the potential of S2ST. Furthermore, recent S2ST research has explored the integration of audio tokens \cite{tjandra2019speech,hsu2021hubert,baevski2020wav2vec,defossez2022high,chung2021w2v, zeghidour22-soundstream} with large language models (LLMs) \cite{wei2023joint,kim2023many,inaguma2022unity,li2023textless,rubenstein2023audiopalm}. This combination offers the potential to improve translation quality and fluency by leveraging the vast linguistic knowledge embedded within LLMs.

\subsection{Real time speech-to-speech translation}
Various architectures have been explored to address the challenges of real-time speech-to-speech translation. A common approach involves cascading automatic speech recognition (ASR), machine translation (MT), and text-to-speech (TTS) systems \cite{sudoh2020simultaneous}. However, this can introduce additional latency. To mitigate latency and improve translation quality, direct S2ST models have been developed. For instance, Ma et al. (2021) proposed a simultaneous S2ST system employing discrete units and a novel variational monotonic multihead attention (V-MMA) mechanism to refine policy learning \cite{ma2021direct}. Similarly, Liu et al. (2021) focused on latency reduction by combining upstream speech translation with incremental text-to-speech (iTTS) \cite{liu2021start}. Non-autoregressive methods have gained traction for their potential speed gains. The TranSpeech architecture \cite{huang2022transpeech} leverages bilateral perturbation, style normalization, and information enhancement for non-autoregressive S2ST, demonstrating significant speed improvements over autoregressive models. Further advancements are evident in the Seamless architecture \cite{barrault2023seamless}, where efficient monotonic multihead attention (EMMA) minimizes latency while maintaining the speaker's vocal style and prosody. Notably, the SimulS2ST architecture \cite{dugan2023learning} offers simultaneous S2ST for a 57 source languages into English, with the flexibility to adjust latency parameters. Importantly, SimulTron establishes a milestone as the first method to demonstrate real-time S2ST on a device.

\section{Model Architecture}

\begin{figure}[t]
    \centering
    \includegraphics[width=1.\columnwidth]{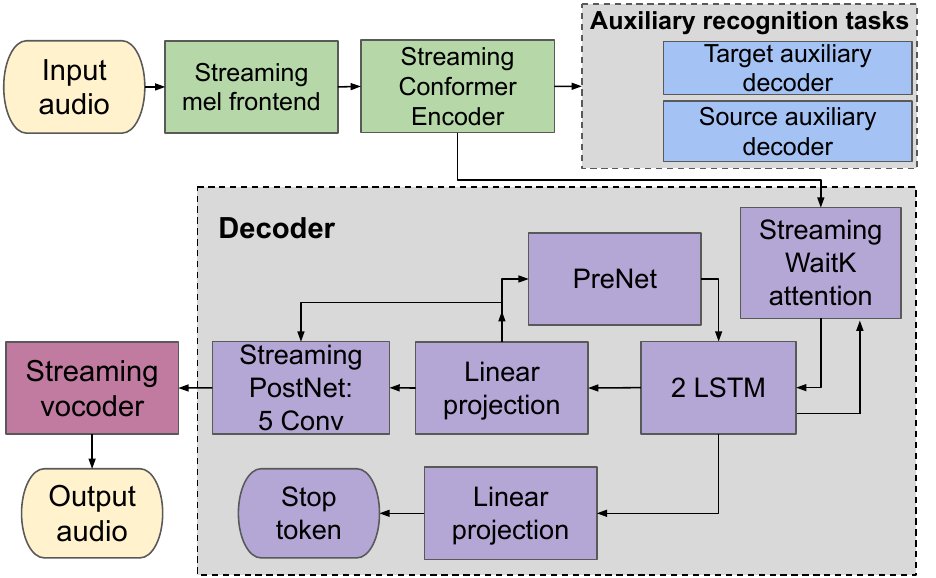}
    \caption{An overview of the proposed SimulTorn architecture. First, the streaming encoder generates a compact representation of the source language input. Subsequently, the decoder, employing wait-k attention, produces a mel-spectrogram representation of the target translation. The MelGAN vocoder then synthesizes the final translated speech output from the mel-spectrogram.}
    \label{fig:architecture}
    \vspace{-4mm}
\end{figure}

The Simultron architecture comprises three models based on the Translatotron 1 architecture tailored for streaming operation: \textbf{Streaming Encoder:} responsible for the real-time encoding of the source language input audio signal. Its design focuses on generating a compact and informative representation of the target language, while maintaining low latency to ensure a responsive system. \textbf{Streaming Decoder:} The autoregressive LSTM decoder uses a wait-k attention to processes the encoded representation from the encoder in sequential fashion, producing acoustic features (mel spectrograms) frame-by-frame. \textbf{Streaming Vocoder:} The final stage in the SimulTron pipeline involves a streaming vocoder that is responsible for converting the generated acoustic features into a time-domain audio waveform. Overview of the proposed SimulTorn architecture is present at Figure~\ref{fig:architecture}.

\subsection{Streaming Encoder}
Input audio frames are processed in streaming mode by mel frontend (Figure~\ref{fig:architecture}) with $80$ mel bins, audio framing is done with frame size $25ms$ and frames step $10ms$. Streaming encoder (Figure~\ref{fig:architecture}) receives two mel frames (every frame with $80$ bins) and process them in streaming mode by $16$ Conformer blocks and one $2$x subsampling layer (because of subsampling layer encoder needs two input frames). Streaming conformer block is composed of a sequence of layers: projection layer followed by causal self attention layer with local context looking into $65$ frames in the past (also called left context), followed by 1d depth-wise convolution layer with kernel size $32$ followed by projection layer and layer normalization. Every layer in the conformer block has residual connection except last layer normalization. Causal self attention layer uses $8$ heads with $256$ feature dimension. All layers in this pipeline: mel frontend, encoder with self attention and convolution layers are causal. Conformer block is open sourced at~\cite{shen2019lingvo}.

\subsection{Streaming Decoder and Vocoder} 

Figure~\ref{fig:architecture} show the spectrogram decoder. It has similar structure with the decoder of the Translatotron model which consists of an attention block, a pre-net, an autoregressive LSTM stack, and PostNet components. We make several changes to it in order to enable causal inference.
We use wait-$k$ attention \cite{ma2018stacl} instead of multi-headed attention. This mechanism allows the decoder to attend to the '$k$' most recently frames of the encoded source sequence, calculating attention weights accordingly. Given the real-time constraints of SimulTron, an initialization step is required where a buffer accumulates '$k$' frames from the encoded source sequence. This accumulation introduces a necessary delay, the duration of which is directly determined by the value of '$k$'.  Importantly, '$k$' represents frames of the encoded source sequence, where each encoded frame encapsulates a $20ms$ segment of the original audio input. Following the wait-k attention, and the LSTM decoder Simultron replaces the PostNet with a causal convolutional PostNet~\cite{rybakov23_interspeech}. Output frames of streaming decoder are sequentially passed to a streaming-capable MelGAN vocoder\cite{rybakov2022real}, which is responsible for synthesizing the final time-domain waveform representation of the audio.

\subsection{Real-time Inference}

To optimize real-time inference and minimize latency, the SimulTron inference architecture employs parallelization between the encoder and decoder components. The system utilizes two distinct threads: one dedicated to the encoder and another for the combined decoder-vocoder module. This concurrent execution strategy enables the encoder to process an impending frame while the decoder and vocoder operate on the current frame. Moreover, alternative model configurations featuring reduced model sizes facilitate the potential for single-threaded execution of both encoder and decoder-vocoder components. The prerequisite for this mode of operation is that the combined computational latency incurred by both components must not exceed the duration of a single output frame. This optimization has significant implications for resource-constrained deployment scenarios. To facilitate real-time inference on resource-constrained mobile devices, SimulTron leverages TFlite (TensorFlow Lite) models \cite{tensorflowTensorFlowLite}. TFlite is a specialized machine learning framework designed for on-device deployment, offering several key advantages. Firstly, TFlite models boast a significantly smaller memory footprint compared to standard TensorFlow models, making them exceptionally well-suited for the memory limitations of mobile environments. Secondly, TFlite models are optimized for accelerated execution, ensuring both low latency and efficient utilization of mobile device hardware. These characteristics are crucial for enabling the smooth and responsive operation of SimulTron in real-world mobile applications.

\section{Experiments and Results}

We conduct experiments utilizing the Conversational dataset \cite{jia2019leveraging} for English-Spanish translation tasks. The model is trained specifically for Spanish-to-English translation. The Conversational dataset offers a substantial 979k utterance pairs, with 1,400 hours of Spanish speech data and 619 hours of English speech data. The Spanish audio is sampled at 16kHz, while the English audio is sampled at 24kHz. For our Spanish-to-English translation experiments, we employed the MuST-C dataset \cite{cattoni2021must}. This dataset comprises 504 hours of English TED Talks audio sampled at 16kHz.  The Spanish component was generated via a proprietary text-to-speech (TTS) system with a female voice and a 24kHz sampling rate. For a full overview of the hyper-parameters used, please refer to Table \ref{tbl:hyperparams}. The Adam optimizer \cite{kingma2014adam} was employed for the training of SimulTron. For audio examples, please refer to the project  \textcolor{blue}{\href{https://enk100.github.io/SimulTron/SimulTron}{website}}. In the mean opinion score (MOS) experiment, we assess the naturalness of the translated speech using a standard 5-point scale with human evaluators. Our latency metric is calculated in the same manner as in iTTS \cite{liu2021start}.

\subsection{Results}

\subsubsection{Conversational dataset}
\textbf{Translation and Acoustics evaluation:}
Table \ref{tab:tab5} compares real-time SimulTron, offline SimulTron, and prior methods on the Conversational Spanish-to-English dataset. The SimulTron real-time model with a delay of $3sec$ ($k=150$) exhibits performance improvement of $0.8$ BLEU point compare to Translatotron 1 despite being a real-time model, attaining a BLEU score of $51.2$ points. This result stems primarily from the shift to a conformer encoder architecture and the use of a $128$ mel spectrogram target representation. Importantly, reducing the initial delay to $k=100$ and $k=50$ (equal to $2sec$ and $1sec$ delay) degrades performance $1.2$ and $3.4$ BLEU points respectively, underscoring the importance of sufficient input context for the decoder's accuracy.  We also evaluate Mean Opinion Scores (MOS) for SimulTron with varying $k$ values ($150$, $100$, $50$) and observe a decline in audio quality with shorter input contexts. Additionally, the performance gap compared to offline models emphasizes the need for high-fidelity streaming vocoder development. We evaluate an offline variant of SimulTron, replacing the streaming conformer encoder with a non-streaming variant and employing a BLSTM decoder with multi-head attention (aligning with Translatotron 1). Offline SimulTron demonstrates a $6.2$ BLEU point improvement over its real-time counterpart, attributed to its ability to leverage full-sentence context at each decoding step, whereas real-time SimulTron is constrained by causal limitations. Moreover, offline SimulTron outperforms Translatotron 1 and 2 by $7.0$ and $1.8$ BLEU points respectively. However, it shows a MOS degradation of $0.57$ points compared to Translatotron 1, which could be addressed with a stronger vocoder. The BLEU gain stems primarily from the shift to a conformer encoder architecture and the use of a $128$ mel spectrogram target representation. While slightly trailing the Cascade approach by $1.4$ BLEU points, this performance remains notable.

\begin{table}[]
\caption{Performance of SimulTron the Conversational Spanish-English dataset and comparison to previous method.}
\label{tab:tab5}
\centering
\setlength{\tabcolsep}{4pt}
\begin{tabular}{llcc}
\toprule
\multicolumn{1}{l}{\textbf{Mode}}  & \textbf{Model}         & \textbf{BLEU ($\uparrow$) } & \textbf{MOS ($\uparrow$) } \\
\midrule
\multirow{5}{*}{\textbf{Offline}}  & Translatotron 1                    & 50.4    &  4.15 ± 0.07                    \\
                                   & Translatotron 2                    & 55.6    & 4.21 ± 0.06              \\
                                   & Cascade (ST $\leftrightarrow$ TTS) & 58.8    & 4.31 ± 0.06                  \\
                                   & Reference (synthetic)             &  81.9    & 3.37 ± 0.09                      \\
                                   & SimulTron                          & 57.4     & 3.58 ± 0.08               \\

\midrule
\multirow{3}{*}{\textbf{Realtime}} & SimulTron (k=50)          & 47.8          & 3.04 ± 0.09       \\
                                   & SimulTron (k=100)         & 50.0          & 3.14 ± 0.09            \\
                                   & SimulTron (k=150)         & 51.2          & 3.35 ± 0.09          \\

\bottomrule
\end{tabular}
\end{table}

\begin{table}[]
\caption{Performance of SimulTron the MuST-C English-Spanish dataset and comparison to previous method.}
\label{tab:tab4}
\centering
\setlength{\tabcolsep}{4pt}
\begin{tabular}{llcc}
\toprule
\multicolumn{1}{l}{\textbf{Mode}}  & \textbf{Model}         & \textbf{BLEU ($\uparrow$) } & \textbf{Latency (s) ($\downarrow$)} \\
\midrule
\multirow{3}{*}{\textbf{Realtime}} 
                                  & iTTS \cite{liu2021start}        & 13.5 & 1.8 \\
                                  & SimulTron (k=125)                & 14.6 & 1.1 \\
                                  & SimulTron (k=150)                & 14.7 & 2.3 \\

\bottomrule
\end{tabular}
\end{table}

\begin{figure}
    \centering
    \includegraphics[width=1.\linewidth]{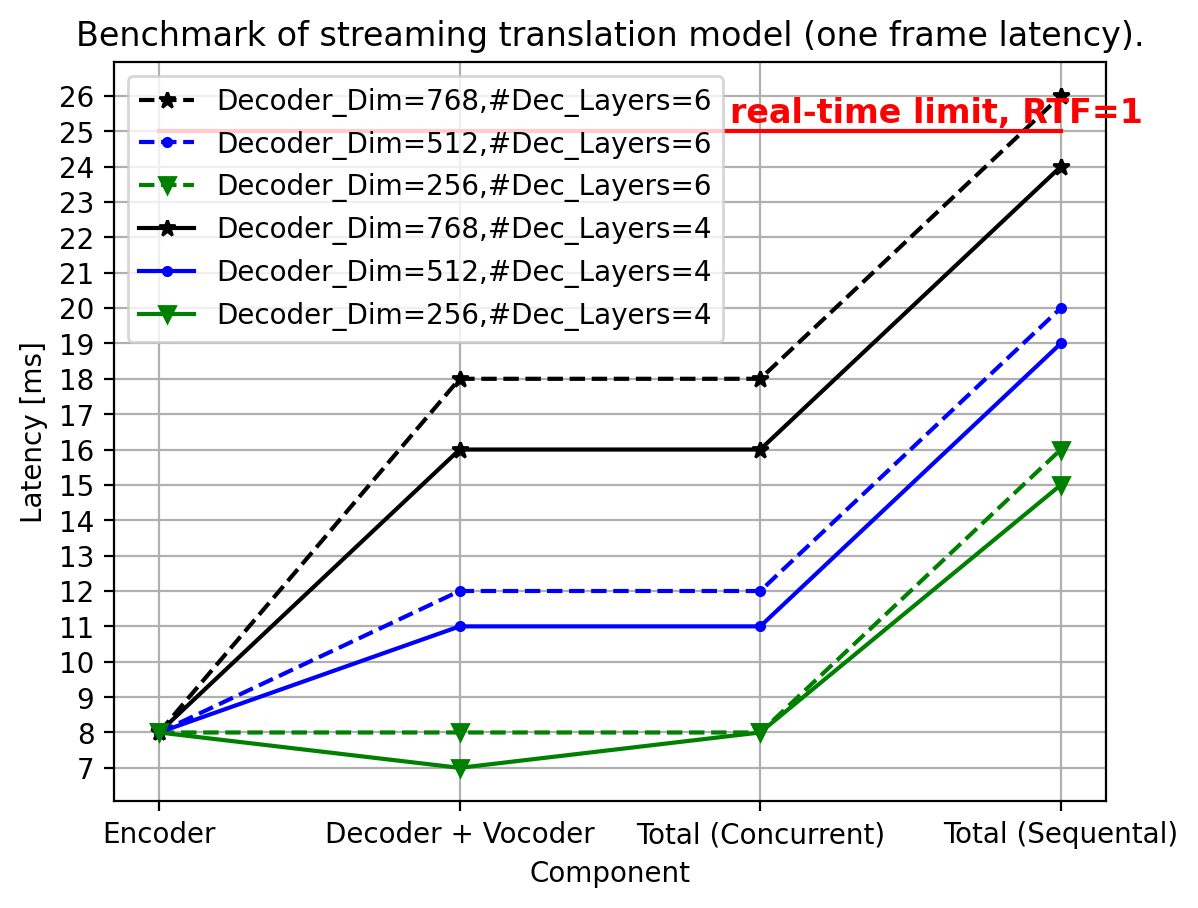}
    \caption{The latency attributed to each model components (Encoder, Decoder+Vocoder), is assessed, with a delineation at the 25-millisecond threshold denoting the practical real-time operational limit.}
    \label{fig:latency1}
\end{figure}

\textbf{Real-time performance evaluation:}
Figure \ref{fig:latency1} presents real-time performance analysis of SimulTron on a Pixel 7 Pro Android device. For streaming translation, the real-time factor (RTF), defined as the ratio of speech output frame length to frame processing time, must exceed $1$. We examine how model configuration and execution mode impact single-frame latency. The x-axis denotes model components (encoder, decoder-vocoder) and their execution mode (concurrent or sequential). Concurrent execution leverages multi-threading, enabling the decoder to process a previous frame while the encoder works on the current one, potentially enhancing efficiency. Sequential execution involves running both components within the same thread. Model variations with differing LSTM decoder dimensionality are represented by colored lines (encoder remains unchanged). As anticipated, models with smaller decoder LSTM dimensions and fewer layers consistently demonstrate lower latencies. Crucially, in concurrent execution, all model configurations exhibit RTFs significantly exceeding $1$, with margins of $7ms$ or more, demonstrating real-time translation. With sequential execution, all but the largest model ($768/6$) also demonstrate feasibility for the tested device.

\textbf{Resource performance:}
Device limitations, particularly memory and storage constraints, are critical considerations for on-device deployment of translation models. Table \ref{tab:tab3} presents memory footprints, RTFs, latencies, and BLEU scores for various SimulTron configurations (differing in LSTM decoder dimension and layer count). As expected, smaller models exhibit reduced memory and storage requirements, making them more suitable for resource-constrained mobile devices. For instance, the $768/6$ model necessitates $562$ Mb of memory and $259$ Mb of storage, whereas the $256/4$ model functions within just $335$ Mb of memory and $148$ Mb of storage. However, this resource efficiency often incurs a trade-off in translation quality, as indicated by slightly lower BLEU scores when the decoder dimension and/or the number of layers are decreased. This highlights the tension between model performance and the feasibility of on-device deployment.

\begin{table}[th]
\caption{Performance summary for SimulTron models.}
\label{tab:tab3}
\centering
\setlength{\tabcolsep}{4pt}
\begin{tabular}{lcccccc}
\toprule
\multicolumn{1}{c}{\begin{tabular}[c]{@{}c@{}} \textbf{Decoder Dim.}\\ \textbf{\# Layer}\end{tabular}} &
  \multicolumn{1}{c}{\begin{tabular}[c]{@{}c@{}} \textbf{768}\\ \textbf{6}\end{tabular}} &
  \multicolumn{1}{c}{\begin{tabular}[c]{@{}c@{}} \textbf{512}\\ \textbf{6}\end{tabular}} &
  \multicolumn{1}{c}{\begin{tabular}[c]{@{}c@{}} \textbf{256}\\ \textbf{6}\end{tabular}} &
  \multicolumn{1}{c}{\begin{tabular}[c]{@{}c@{}} \textbf{768}\\ \textbf{4}\end{tabular}} &
  \multicolumn{1}{c}{\begin{tabular}[c]{@{}c@{}} \textbf{512}\\ \textbf{4}\end{tabular}} &
  \multicolumn{1}{c}{\begin{tabular}[c]{@{}c@{}} \textbf{256}\\ \textbf{4}\end{tabular}} \\
 \midrule
Latency [ms]($\downarrow$) & 18 & 12 & 8 & 16 & 11 & 8 \\
RTF ($\uparrow$) & 1.4x & 2x & 3.1x & 1.5x  & 2.3x & 3.1x \\
Size [MB]($\downarrow$) & 259 & 195 & 154 & 217  & 175 & 148 \\
Memory [MB]($\downarrow$) & 562 & 431 & 347 & 476  & 391 & 335\\
BLEU ($\uparrow$) & 51.2 & 50.9 & 49.8 & 51.0  & 50.7 & 49.6\\
\bottomrule
\end{tabular}
\end{table}

\textbf{Post-Training Quantization:}
We implement post-training dynamic range quantization to reduce model size and improve efficiency. As can be seen in Table \ref{tab:tab1} weights that are quantized to int8, dramatically decreasing latency by factor of $2$ to the encoder, $2$ and $1.8$ for the decoder with dimension of $1024$ and $768$ respectively. For model storage size dramatically decreasing by factor of $3.7$ to the encoder, $3.4$ and $3.1$ for the decoder with dimension of 1024 and 768 respectively. Moreover, memory footprints dramatically decreasing by factor of $4.7$ to the encoder, $4.2$ and $3.5$ for the decoder with dimension of $1024$ and $768$ respectively. 
\begin{table}[th]
\caption{Latency in sec. 10s benchmark of non-streaming SimulTron on Pixel7pro using float32 and Int8 tensor operations. Lower is better.}
\label{tab:tab1}
\centering
\setlength{\tabcolsep}{4pt}
\begin{tabular}{lccccccc}
\toprule
\multicolumn{1}{l}{\textbf{Metric}} & \multicolumn{1}{c}{\textbf{Type}} &
  \multicolumn{1}{c}{\textbf{Encoder}}  & \multicolumn{1}{c}{\begin{tabular}[c]{@{}c@{}} \textbf{Decoder} \\ \textbf{1024}\end{tabular}} & \multicolumn{1}{c}{\begin{tabular}[c]{@{}c@{}} \textbf{Decoder} \\ \textbf{768}\end{tabular}} \\ 
\midrule
\multicolumn{1}{c}{\textbf{Latency [ms]}}  & Float32 & 8  & 26 & 18  \\
\multicolumn{1}{c}{{($\downarrow$)}}   & Int8 & 4  & 13 & 10 \\
\midrule
\multicolumn{1}{c}{\textbf{Size [MB]}}  & Float32 & 104 & 241 & 155 \\
\multicolumn{1}{c}{{($\downarrow$)}} & Int8 & 28 & 71 & 50 \\   
\midrule
\multicolumn{1}{c}{\textbf{Memory [MB]}}  & Float32 & 227 & 509 & 335 \\
\multicolumn{1}{c}{{($\downarrow$)}}  & Int8 & 48 & 119 & 94 \\  

\bottomrule
\end{tabular}
\end{table}

\begin{table}[t]
\setlength{\tabcolsep}{0.5em}
\centering
\caption{Table of hyper-parameters.}
\setlength{\tabcolsep}{4pt}
\begin{tabular}{l@{\hspace{0.2em}}rr}
    \toprule
    \textbf{Hyper-parameter}  &  & \\
    \midrule
    Input / output sample rate (Hz) & 16k / 24k \\
    Learning rate  & 0.001 \\
    Batch Size & 1024  \\
    Encoder (layers $\times$ dim)  & 16$\times$256  \\
    Decoder (layers $\times$ dim)  & 6$\times$678  \\
    Decoder Attn Wait-K  (Input frames) & 150 \\
    Decoder Attn dropout prob & 0.1 \\
    Decoder output mel channels & 128  \\
    Encoder frame size [ms] & 20  \\
    Decoder frame size [ms] & 25  \\
    Vocoder input dim & 128  \\
    Vocoder mel edges Hz  & 20Hz-12kHz \\
    \bottomrule
\end{tabular}
\label{tbl:hyperparams}
\end{table}

\subsubsection{MuST-C dataset}

Table \ref{tab:tab4} highlights the comparative analysis of our method against a leading real-time S2ST approach. Our method improves the iTTS method \cite{liu2021start} by 1.1 and 1.2 points for k=125 and k=150, respectively. Importantly, we also demonstrate a latency reduction of 0.7 seconds compared to the iTTS model when k=125.

\section{Conclusion}
SimulTron, a real-time S2ST architecture designed to provide real-time, on-device simultaneous speech-to-speech translation. SimulTron builds upon the Translatotron framework, incorporating optimizations like a causal conformer encoder, wait-k attention, a convolutional post-net, and a streaming vocoder that enable efficient, streaming S2ST operation. SimulTron surpasses Translatotron 2 in offline evaluations and maintains comparable performance to Translatotron 1 in real-time Spanish-to-English translation settings. Additionally, SimulTron achieves better BLEU scores and latency compared to the iTTS real-time S2ST approach on the MuST-C dataset. Moreover, SimulTron's careful design enables it to function within the constraints of mobile devices. By bringing real-time, simultaneous translation directly to mobile devices, we envision a future where language barriers are significantly reduced, fostering greater understanding and collaboration across cultures. Future work will focus on expanding SimulTron's multilingual capabilities, further optimizing for various mobile hardware, and exploring techniques to enhance translation quality under challenging acoustic conditions.

\clearpage

\bibliographystyle{IEEEtran}
\bibliography{mybib}

\begin{thebibliography}{10}
\providecommand{\url}[1]{#1}
\csname url@samestyle\endcsname
\providecommand{\newblock}{\relax}
\providecommand{\bibinfo}[2]{#2}
\providecommand{\BIBentrySTDinterwordspacing}{\spaceskip=0pt\relax}
\providecommand{\BIBentryALTinterwordstretchfactor}{4}
\providecommand{\BIBentryALTinterwordspacing}{\spaceskip=\fontdimen2\font plus
\BIBentryALTinterwordstretchfactor\fontdimen3\font minus \fontdimen4\font\relax}
\providecommand{\BIBforeignlanguage}[2]{{%
\expandafter\ifx\csname l@#1\endcsname\relax
\typeout{** WARNING: IEEEtran.bst: No hyphenation pattern has been}%
\typeout{** loaded for the language `#1'. Using the pattern for}%
\typeout{** the default language instead.}%
\else
\language=\csname l@#1\endcsname
\fi
#2}}
\providecommand{\BIBdecl}{\relax}
\BIBdecl

\bibitem{jia2019direct}
Y.~Jia~et al, ``Direct speech-to-speech translation with a sequence-to-sequence model,'' \emph{Proc. Interspeech}, pp. 1123--1127, 2019.

\bibitem{zhang2021uwspeech}
C.~Zhang~et al, ``Uwspeech: Speech to speech translation for unwritten languages,'' in \emph{Proc. AAAI}, vol.~35, 2021, pp. 14\,319--14\,327.

\bibitem{lee2021direct}
A.~Lee~et al, ``Direct speech-to-speech translation with discrete units,'' \emph{arXiv:2107.05604}, 2021.

\bibitem{lee2021textless}
------, ``Textless speech-to-speech translation on real data,'' \emph{arXiv:2112.08352}, 2021.

\bibitem{huang2022transpeech}
R.~Huang~et al, ``{TranSpeech}: Speech-to-speech translation with bilateral perturbation,'' \emph{arXiv:2205.12523}, 2022.

\bibitem{sudoh2020simultaneous}
K.~Sudoh, T.~Kano, S.~Novitasari, T.~Yanagita, S.~Sakti, and S.~Nakamura, ``Simultaneous speech-to-speech translation system with neural incremental asr, mt, and tts,'' \emph{arXiv preprint arXiv:2011.04845}, 2020.

\bibitem{ma2021direct}
X.~Ma, H.~Gong, D.~Liu, A.~Lee, Y.~Tang, P.-J. Chen, W.-N. Hsu, P.~Koehn, and J.~Pino, ``Direct simultaneous speech-to-speech translation with variational monotonic multihead attention,'' \emph{arXiv preprint arXiv:2110.08250}, 2021.

\bibitem{liu2021start}
D.~Liu, C.~Wang, H.~Gong, X.~Ma, Y.~Tang, and J.~Pino, ``From start to finish: Latency reduction strategies for incremental speech synthesis in simultaneous speech-to-speech translation,'' \emph{arXiv preprint arXiv:2110.08214}, 2021.

\bibitem{barrault2023seamless}
L.~Barrault, Y.-A. Chung, M.~C. Meglioli, D.~Dale, N.~Dong, M.~Duppenthaler, P.-A. Duquenne, B.~Ellis, H.~Elsahar, J.~Haaheim \emph{et~al.}, ``Seamless: Multilingual expressive and streaming speech translation,'' \emph{arXiv preprint arXiv:2312.05187}, 2023.

\bibitem{dugan2023learning}
L.~Dugan, A.~Wadhawan, K.~Spence, C.~Callison-Burch, M.~McGuire, and V.~Zordan, ``Learning when to speak: Latency and quality trade-offs for simultaneous speech-to-speech translation with offline models,'' \emph{arXiv preprint arXiv:2306.01201}, 2023.

\bibitem{nakamura2006atr}
S.~Nakamura~et al, ``The {ATR} multilingual speech-to-speech translation system,'' \emph{IEEE Trans. ASLP}, vol.~14, no.~2, pp. 365--376, 2006.

\bibitem{wahlster2013verbmobil}
W.~Wahlster, \emph{Verbmobil: Foundations of speech-to-speech translation}.\hskip 1em plus 0.5em minus 0.4em\relax Springer Science \& Business Media, 2013.

\bibitem{lavie1997janus}
A.~Lavie~et al, ``{JANUS-III}: Speech-to-speech translation in multiple languages,'' in \emph{Proc. ICASSP}, vol.~1, 1997, pp. 99--102.

\bibitem{jia2022translatotron}
Y.~Jia~et al, ``Translatotron 2: High-quality direct speech-to-speech translation with voice preservation,'' in \emph{Proc. ICML}, 2022, pp. 10\,120--10\,134.

\bibitem{nachmani2023translatotron}
E.~Nachmani, A.~Levkovitch, Y.~Ding, C.~Asawaroengchai, H.~Zen, and M.~T. Ramanovich, ``Translatotron 3: Speech to speech translation with monolingual data,'' \emph{arXiv preprint arXiv:2305.17547}, 2023.

\bibitem{wang2022simple}
C.~Wang~et al, ``Simple and effective unsupervised speech translation,'' \emph{arXiv:2210.10191}, 2022.

\bibitem{fu2023improving}
Y.-K. Fu, L.-H. Tseng, J.~Shi, C.-A. Li, T.-Y. Hsu, S.~Watanabe, and H.-y. Lee, ``Improving cascaded unsupervised speech translation with denoising back-translation,'' \emph{arXiv preprint arXiv:2305.07455}, 2023.

\bibitem{tjandra2019speech}
A.~Tjandra, S.~Sakti, and S.~Nakamura, ``Speech-to-speech translation between untranscribed unknown languages,'' in \emph{2019 IEEE Automatic Speech Recognition and Understanding Workshop (ASRU)}.\hskip 1em plus 0.5em minus 0.4em\relax IEEE, 2019, pp. 593--600.

\bibitem{hsu2021hubert}
W.-N. Hsu~et al, ``Hubert: Self-supervised speech representation learning by masked prediction of hidden units,'' \emph{IEEE/ACM Transactions on Audio, Speech, and Language Processing}, vol.~29, pp. 3451--3460, 2021.

\bibitem{baevski2020wav2vec}
A.~Baevski~et al, ``wav2vec 2.0: A framework for self-supervised learning of speech representations,'' \emph{Advances in neural information processing systems}, vol.~33, pp. 12\,449--12\,460, 2020.

\bibitem{defossez2022high}
A.~D{\'e}fossez~et al, ``High fidelity neural audio compression,'' \emph{arXiv preprint arXiv:2210.13438}, 2022.

\bibitem{chung2021w2v}
Y.-A. Chung~et al, ``W2v-bert: Combining contrastive learning and masked language modeling for self-supervised speech pre-training,'' in \emph{2021 IEEE Automatic Speech Recognition and Understanding Workshop (ASRU)}.\hskip 1em plus 0.5em minus 0.4em\relax IEEE, 2021, pp. 244--250.

\bibitem{zeghidour22-soundstream}
N.~Zeghidour, A.~Luebs, A.~Omran, J.~Skoglund, and M.~Tagliasacchi, ``Soundstream: An end-to-end neural audio codec,'' \emph{{IEEE} {ACM} Trans. Audio Speech Lang. Process.}, vol.~30, pp. 495--507, 2022.

\bibitem{wei2023joint}
K.~Wei~et al, ``Joint pre-training with speech and bilingual text for direct speech to speech translation,'' in \emph{ICASSP 2023-2023 IEEE International Conference on Acoustics, Speech and Signal Processing (ICASSP)}.\hskip 1em plus 0.5em minus 0.4em\relax IEEE, 2023, pp. 1--5.

\bibitem{kim2023many}
M.~Kim~et al, ``Many-to-many spoken language translation via unified speech and text representation learning with unit-to-unit translation,'' \emph{arXiv preprint arXiv:2308.01831}, 2023.

\bibitem{inaguma2022unity}
H.~Inaguma~et al, ``Unity: Two-pass direct speech-to-speech translation with discrete units,'' \emph{arXiv preprint arXiv:2212.08055}, 2022.

\bibitem{li2023textless}
X.~Li~et al, ``Textless direct speech-to-speech translation with discrete speech representation,'' in \emph{ICASSP 2023-2023 IEEE International Conference on Acoustics, Speech and Signal Processing (ICASSP)}.\hskip 1em plus 0.5em minus 0.4em\relax IEEE, 2023, pp. 1--5.

\bibitem{rubenstein2023audiopalm}
P.~K. Rubenstein~et al, ``Audiopalm: A large language model that can speak and listen,'' \emph{arXiv preprint arXiv:2306.12925}, 2023.

\bibitem{shen2019lingvo}
J.~Shen, P.~Nguyen, Y.~Wu, Z.~Chen \emph{et~al.}, ``Lingvo: a modular and scalable framework for sequence-to-sequence modeling,'' 2019.

\bibitem{ma2018stacl}
M.~Ma, L.~Huang, H.~Xiong, R.~Zheng, K.~Liu, B.~Zheng, C.~Zhang, Z.~He, H.~Liu, X.~Li \emph{et~al.}, ``Stacl: Simultaneous translation with implicit anticipation and controllable latency using prefix-to-prefix framework,'' \emph{arXiv preprint arXiv:1810.08398}, 2018.

\bibitem{rybakov23_interspeech}
O.~Rybakov, F.~Biadsy, X.~Zhang, L.~Jiang, P.~Meadowlark, and S.~Agrawal, ``{Streaming Parrotron for on-device speech-to-speech conversion},'' in \emph{Proc. INTERSPEECH 2023}, 2023, pp. 2033--2037.

\bibitem{rybakov2022real}
O.~Rybakov, M.~Tagliasacchi, Y.~Li, L.~Jiang, X.~Zhang, and F.~Biadsy, ``{Real time spectrogram inversion on mobile phone},'' in \emph{Proc. INTERSPEECH 2023}, 2023, pp. 4314--4318.

\bibitem{tensorflowTensorFlowLite}
``{T}ensor{F}low {L}ite | {M}{L} for {M}obile and {E}dge {D}evices --- tensorflow.org,'' \url{https://www.tensorflow.org/lite}, [Accessed 05-03-2024].

\bibitem{jia2019leveraging}
Y.~Jia~et al, ``Leveraging weakly supervised data to improve end-to-end speech-to-text translation,'' in \emph{Proc. ICASSP}, 2019, pp. 7180--7184.

\bibitem{cattoni2021must}
R.~Cattoni, M.~A. Di~Gangi, L.~Bentivogli, M.~Negri, and M.~Turchi, ``Must-c: A multilingual corpus for end-to-end speech translation,'' \emph{Computer Speech \& Language}, vol.~66, p. 101155, 2021.

\bibitem{kingma2014adam}
D.~P. Kingma and J.~Ba, ``Adam: A method for stochastic optimization,'' \emph{arXiv preprint arXiv:1412.6980}, 2014.

\end{thebibliography}

\end{document}